\documentclass[%
 reprint,
 amsmath,amssymb,
 aps,
]{revtex4-2}

\usepackage{graphicx}
\usepackage{dcolumn}
\usepackage{bm}
\usepackage{lipsum}

\begin{document}

\preprint{APS/123-QED}

\title{Multiple states of turbulence at vanishing inertia}

\author{Ziyin Lu}
\author{Bj\"orn Hof}%
 \email{bhof@ist.ac.at}
\affiliation{%
Institute of Science and Technology Austria, 3400 Klosterneuburg, Austria}%

\date{\today}

\begin{abstract}
Based on everyday experience fluid flows tend to be ordered and quiescent if inertial forces are low and held in check by viscosity. This intuition spectacularly fails in the case of complex macromolecular fluids like polymer melts, paints and biofluids. In such cases elastic fluid properties can drive turbulent motions at moderate and even vanishing Reynolds numbers. By studying viscoelastic flows in curved pipes we demonstrate that this low inertia phenomenology results from the competition of two hydrodynamic instabilities and respectively from the co-existence and interdependence of two distinct turbulent states. Unexpectedly the established categories of elastic and elasto-inertial turbulence (ET and EIT) fail to demarcate the actual turbulent states, fundamentally changing the perception of this phenomenon a century after its discovery.

\end{abstract}

\maketitle
Turbulence fundamentally alters the motion of fluids and determines the transport of mass, momentum and heat. While the onset of turbulence is well understood for ordinary fluids like air and water, its occurrence in complex, viscoelastic fluids such as polymeric liquids, colloids and gels follows drastically different rules. The common diagnostic of turbulence, i.e. the magnitude of the Reynolds number, a measure of inertial to viscous forces, is of little relevance here. Rather than inertia, the fluid's elasticity is responsible for the nonlinear interactions that permit turbulence to arise.

The unexpected occurrence of disordered motions at low inertia can be disruptive in applications such as material processing \cite{carrillo2014gelatinized,avila2015mechanisms,bertola2003experimental}), yet at the same time it offers opportunities for mixing and heat transport in microfluidics, and hence in situations where ordinarily turbulence enhanced mixing would be unavailable.
In these situations polymeric additives can be employed to functionalize liquids and trigger elastic instabilities\cite{Groisman-2001-Nature}. Rather than the Reynolds number ($\mathit{Re}=Ud/\nu$, where $U$ and $d$ are characteristic velocity and length scales of the flow and $\nu$ the fluid's kinematic viscosity), the relevant control parameter that decides the flow's stability is now the Weissenberg number, which accordingly measures the ratio of elastic to viscous forces ($\mathit{Wi}=\lambda {\dot \gamma}$, where $\lambda$ is the polymer relaxation time and ${\dot \gamma}$ the shear rate).
In correspondence with these two control parameters, three types of turbulence are commonly distinguished for viscoelastic fluids:  purely elastic turbulence, ET,  at large $\mathit{Wi}$ and negligible $\mathit{Re}$, elasto-inertial turbulence, EIT, if both parameters are relevant, and ordinary inertial turbulence, IT, if $\mathit{Re}$ is large and $\mathit{Wi}$ negligible. 

ET was originally discovered in curved streamline flows (cylindrical Couette flows \cite{larson1990purely}, plate and cone rheometers \cite{Groisman-2000-Nature} etc.) of polymer solutions where the combination of elasticity and geometry gives rise to hoop stresses \cite{Shaqfeh-1996-ANRFM} that render the laminar flow linearly unstable and drive the chaotic flow state. 

Conversely, this known linear instability is absent in rectilinear geometries, such as straight pipes and channels. Yet observations of elastically driven chaotic motions in pipe flow date back to experiments in colloidal suspensions a century ago \cite{Ostwald-1926-KOLL}. Studies of polymer solutions in pipe flow investigated the onset of this instability, its competition with inertial turbulence and its possible relevance to polymer drag reduction, specifically to what is known as the maximum drag reduction asymptote \cite{samanta2013, choueiri2021}. Given that these earlier observations in pipe flow tend to be at finite inertia, these viscoelastic turbulent motions are referred to as elasto-inertial turbulence.   
Recent theoretical investigations discovered that viscoelastic pipe \cite{Garg-2018-PRL} and channel flows \cite{Khalid-2021-JFM} are linearly unstable to a center mode traveling wave, which is believed to support these elasto-inertia motions. It is noteworthy that, instead of a center mode, the initial investigation of EIT  reported structures associated with wall modes, i.e. near wall vortices and sheets of polymer stretch.   

The center mode instability characteristically ceases to exist at some minimum inertia level \cite{Garg-2018-PRL}, which seemingly confirms the elasto-inertial nature of this instability. However, for channel flows the center mode has been found to, either supercritically \cite{Khalid-2021-JFM, Khalid-2021-PRL} or subcritically \cite{Page-2020-PRL, Buza-2022-JFM-2}, extend to zero inertia. The approach to the zero inertia limit in rectilinear flows led to speculation that EIT and ET have the same origin\cite{lellep2024purely}, inferring a unique state of turbulence at zero inertia.

Unfolding the parameter space to curved pipes we demonstrate that the state of elastic turbulence is not unique. While at low curvature the center mode instability gives rise to a weakly fluctuating state, at high curvature elastic turbulence is driven by hoop stresses and exhibits order of magnitude larger fluctuation levels. Surprisingly the hoop stress driven state persists to the weakly curved and straight pipe case, where, enabled by the primary center mode, it sets in as a secondary instability and subsequently dominates the dynamics. The near wall sheet like structures that characterize the eventual state of E(I)T are shown to originate from this hoop stress instability. 

\vspace{1cm}
Initial experiments are carried out in a $d=4$ mm diameter straight pipe (consisting of a 60 cm semi-flexible polyurethane tube) of aspect ratio $L/d = 150$. The working fluid is a 200 part-per-million (ppm) solution of polyacrylamide (PAAM, molecular weight of 18 million Dalton) in a Newtonian solvent composed of 15\% water and 85\% glycerol. From the inlet, the flow is left to evolve across a streamwise distance of $100 d$.  Downstream of this location a differential pressure sensor is connected via two 0.5 mm holes in the pipe wall (spaced by 10d in the streamwise direction). Monitoring the flow's fluctuation levels the sensor allows to readily distinguish laminar from turbulent motions\cite{samanta2013,choueiri2021}  and to determine the onset of turbulence.  
Figure 1(a) shows the evolution of the pressure fluctuations with increasing Reynolds number. Fluctuations in this case arise at $\mathit{Re}\approx 1.5$. The onset is non-hysteretic and is found to occur at the same $\mathit{Re}$, irrespective if the upstream flow is perturbed \cite{samanta2013,choueiri2021} (e.g. by injection of a jet or insertion of an obstacle). 

To investigate the flow structure, in addition, velocity fields were measured using particle image velocimetry, PIV. For this purpose a small amount of micron sized particles are added to the fluid, and the flow is illuminated with a laser in a cross-sectional (radial-streamwise) plane. Close to onset, the fluctuation levels are largest in the vicinity of the pipe center (see blue curve in Fig. 1b) in accordance with the  center mode instability\cite{Garg-2018-PRL} and earlier experiments\cite{choueiri2021}, and decrease towards the wall (Fig.1b). Also in line with previous experiments\cite{choueiri2021}, the nature of the flow is found to change with increasing $\mathit{Re}$ from a center to a wall mode (Fig.1b) at what appears to be a secondary instability\cite{choueiri2021}. As shown in Fig. 1(a), this transformation occurs at $\mathit{Re} \approx7$ and is signified by a sharp increase in fluctuation levels. 

In order to explore the nature and origin of the wall mode, we next investigate the role of streamline curvature by coiling the pipe around a cylindrical inner core, resulting in a  helical pipe of a chosen curvature ratio of $d/R$ and a pitch corresponding to the pipe outer diameter, 6 mm. 
As shown in Fig. 1(c), the onset threshold of the center mode instability (black circles) is unaffected by curvature. The wall mode (red circles) on the other hand is found to be curvature dependent and the threshold decreases with increasing $d/R$. Eventually, for $d/R\approx0.06$ it becomes the primary instability and the threshold monotonically tends to zero as the curvature further increases. 
Such monotonic curvature dependence is well known from other curvilinear flows for the hoop stress mode. The correspondence with the hoop stress mode is further confirmed by fitting the Pakdel-McKinley\cite{pakdel1996elastic} criterion to our data. As shown in inset of Fig. 1(c), the curvature dependence of the critical Weissenberg number is nicely captured by this criterion. The data does not only follow the Pakdel-McKinley scaling at high curvature (low R/d), but the scaling surprisingly persists to lower curvatures where the hoop stress instability is preceded by the center mode instability. 

We next test the robustness of this scenario and check if this intricate interplay between instabilities persists to the inertialess regime. To do so, we increase the fluid elasticity, which is quantified by the elasticity number $\mathit{E}=\mathit{Wi}/\mathit{Re}=\lambda {\nu}/d^2$. To increase $\mathit{E}$, the pipe diameter was decreased to $d = 1.6$ mm while a slightly lower glycerol and polymer concentration was chosen (80\% glycerol, 50 ppm PAAM) in order to reach the required flow rates, given the higher resistance in the smaller tube diameter. Otherwise the set up was in all aspects equivalent to the 4 mm pipe (see Materials and Methods for details). 

As shown in Fig. 2(a), at this higher Elasticity number of $E \approx 80$, the transition threshold of the center mode has dropped to $\mathit{Re} = 7\times10^{-2}$ and inertia is hence negligible.  For the low curvature cases (black circles), the onset again occurs at a constant Reynolds number. The hoop stress instability as expected is the primary instability at high curvature where it continues to zero inertia. Also here the hoop stress instability is not limited to high curvatures but it persists as a secondary instability to the straight pipe case ($d/R = 0$). 

In experiments the two instabilities can be readily distinguished from the turbulent fluctuation levels close to onset. As shown in figure 2B the center mode (blue symbols) irrespective of curvature only exhibits very low fluctuation levels. For the hoop stress mode (brown, orange, yellow symbols) on the other hand the fluctuations increase much faster with distance from the critical point. The rate of increase is approximately constant, irrespective if the hoop stress mode sets in as a primary or secondary instability. 
Simultaneously measured velocity profiles, more precisely the velocity fluctuation profiles, Fig. 2(c), show that the center mode keeps the central peak also at finite curvature. The hoop stress mode (yellow to red) on the other hand exhibits the largest fluctuation near the wall. 

The distinction between the two modes is equally confirmed by the velocity fields, where as shown in Fig. 3(a) the center mode with its characteristic chevron type structure is found for the straight and mildly curved cases. In contrast the flow field of the hoop stress mode (panel b) is characterized  by order of magnitude stronger streaks. The identical structural composition is found regardless if the hoop stress mode is the primary or secondary instability.  As a final distinction, we compare the power spectra of the flow fields in the center mode regime to that in the hoop stress regime (Fig. 3c). The spectra of the latter have a slope of $\leq -3$, in accordance with the prediction for purely elastic turbulence arising from the hoop stress mode\cite{Groisman-2000-Nature,fouxon2003spectra}. In contrast, power spectra of the center mode flow field, at least for the frequencies that can be resolved in experiments, are much shallower and have slopes of $\approx-2$ . 

A remaining question is how the hoop stress instability can persist to the straight pipe and here earlier studies had proposed that finite amplitude perturbations could provide the required streamline curvature\cite{bertola2003experimental,Morozov-2005-PRL,Morozov-2007-PR}. While in laminar pipe flow curved streamlines are absent, this does not apply to the flow fields subsequent to the onset of the center mode . We hence propose that the center mode, by curving the streamlines, eventually promotes the hoop stress mode.

Finally, we would like to point out various seemingly conflicting past observations regarding the nature of viscoelastic pipe and channel flows. Initial studies of EIT\cite{samanta2013,Dubief-2013-POF} characterized this state based on the near wall sheets of polymer stretch whereas subsequent studies associated it with center mode structures. These fundamental differences in flow structure are finally resolved by our study, clarifying that what was considered as the state of EIT are in reality two different chaotic dynamical states arising from their respective instabilities. 
It becomes apparent from the results of the present study that the current naming convention and the categories of ET and EIT based on levels of inertia and elasticity do not capture the actual dynamical states. The hoop stress driven turbulent motions, featuring wall modes, arise at vanishing inertia in sufficiently curved tubes (Fig. 3b top panel) and consequently are categorized as ET in this case. In the straight pipe (Fig. 3b bottom panel), the identical dynamical state arises at finite Re and is now referred to as EIT. At the same time the state of hoop stress turbulence in Fig. 3(b) (top panel) and the center mode turbulence of Fig. 3(a) both arise at $Re<10^{-1}$ and despite being fundamentally different in nature would both classify as ET. 

It is noteworthy that also in the Newtonian context 'turbulence' often refers to flows of different nature which in turn are associated with distinct instabilities. Individual driving mechanisms, e.g. body forces, give rise to chaotic motions composed of case-specific coherent structures and sustaining mechanisms.  One example is the earth's liquid core, i.e. the geodynamo, where turbulent motions are driven by Lorentz and Coriolis forces, while inertia is irrelevant. 
What sets the present case apart, is that with elasticity a single driving force is responsible for two distinct turbulent states. This unusual circumstance of two types of turbulent motion being governed by the identical control parameter (Wi), is responsible for the exceptionally rich dynamics encountered in viscoelastic flows and the conceptual challenges associated with this phenomenon.

\begin{figure*}
    \centering
    \includegraphics[width=1.0\linewidth]{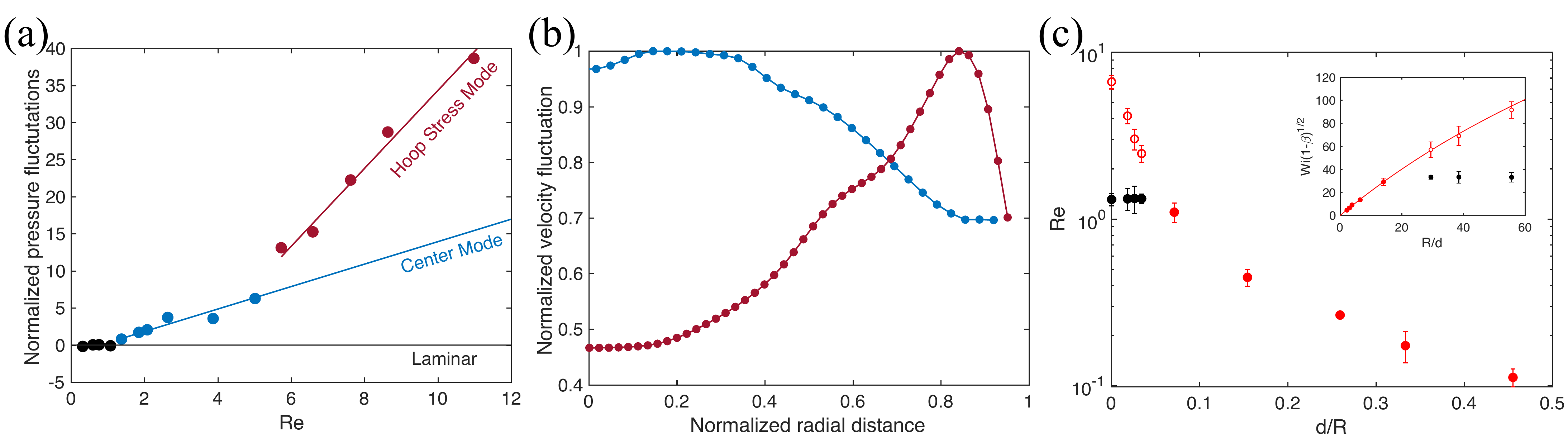}
    \caption{Onset of multiple turbulent states and transformation from a center to a wall mode.
		(a) Normalized pressure fluctuation levels as a function of Reynolds number used to determine the onset. The fluctuation levels are calculated by $(\sigma_p-\sigma_{p}^{lam})/\sigma_{p}^{lam}$, where $\sigma_p$ is the standard deviation of pressure signals of each data point and $\sigma_{p}^{lam}$ is that of points representing laminar flow. Straight lines are fitted to different states for illustration. (b) Streamwise velocity fluctuations, $u^\prime/u^\prime_{\mathit{max}}$, as a function of distance from the pipe center, $r/(d/2)$, obtained from PIV. At $\mathit{Re} = 50, \mathit{Wi} = 50$, the peak of the fluctuations profile appears close to the pipe center. At $\mathit{Re} = 50, \mathit{Wi} = 75$, the peak has shifted towards the wall. (c) The onset of the center mode (black circles), the hoop stress mode as a primary instability (red circles), and the hoop stress as a secondary instability (open circles) as a function of curvature ratio, $d/R$. Inset depicts critical $\mathit{Wi}$ as a function of inverse curvature ratio, $R/d$. The red line is the Pakdel-McKinley criterion of two scales, written in the form of $\mathit{Wi}_c (1-\beta)^{1/2} = M_c[a/(R/d)+b]^{-1}$ (where $\beta$ denotes viscosity ratio between solvent and solution, and $M_c = 4.08$, $a = 1.83$ and $b = 0.01$ are fitting parameters), fitted to the data points of primary hoop stress mode (red solid circles). Each data point is obtained from 3 to 10 rehearsals of the experiment shown in (A). Error bars indicate standard deviation.}
    \label{fig:fig1}
\end{figure*}

\begin{figure*}
    \centering
    \includegraphics[width=1.0\linewidth]{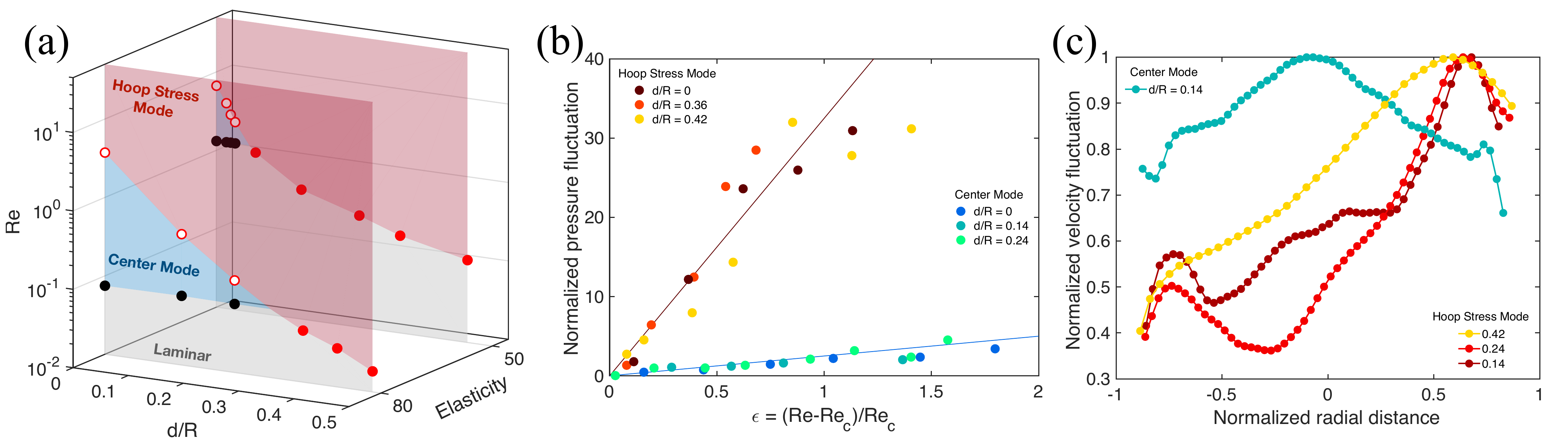}
    \caption{Multiple turbulent states and the approach to the inertialess regime.
		(a) Onset of the center and hoop stress modes as a function of curvature ratio for 85\% glycerol-200 ppm PAAM in 4 mm pipe ($\mathit{E} \approx 50$) and 80\% glycerol-50 ppm PAAM in 1.6 mm pipe ($\mathit{E} \approx 80$). (b) Pressure fluctuation levels as a function of distance from the critical points of the center and hoop stress modes, respectively, in 1.6 mm pipes. Profiles of the hoop stress mode are shifted vertically by the amount of normalized pressure fluctuation at the onset. Solid lines are guides to the eye. (c) Streamwise velocity fluctuations, $u^\prime/u^\prime_{\mathit{max}}$, as a function of distance from the pipe center, $r/(d/2) = 0$, measured using PIV in 1.6 mm curved pipes. The pipe walls concave left. Data points in the vicinity of walls are omitted because of large uncertainties.}
    \label{fig:fig2}
\end{figure*}

\begin{figure*}
    \centering
    \includegraphics[width=1.0\linewidth]{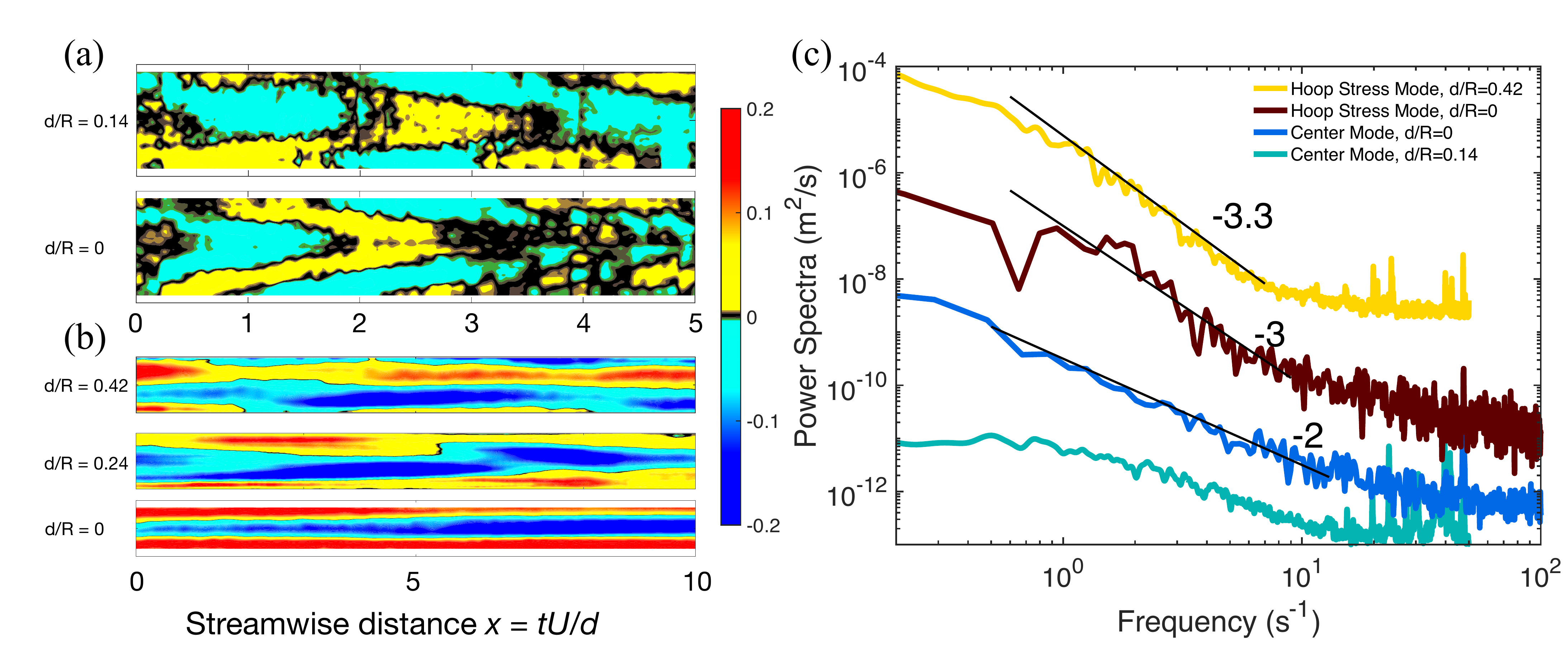}
    \caption{Distinction between turbulent states.
		(a and b) Streamwise velocity fluctuations, $u^\prime/U$, close to the onsets of the center mode and wall mode, respectively. For the center mode, weak fluctuations (represented by yellow for positive $u^\prime$ regions and cyan for negative ones) are spatially organized in a chevron pattern, while for the wall mode, strong fluctuations (red and blue) emerge as streaks elongated in the streamwise direction. (c) Power spectra of the streamwise velocity component for turbulent states corresponding to the center and hoop stress modes. The velocity signals are sampled at locations close to the center line of the pipe. }
    \label{fig:fig3}
\end{figure*}

\subsection{Acknowledgments}
We gratefully acknowledge helpful discussions and feedback by colleagues during the EFDC1 (Aachen, September 2024), the Chaotic Flows in Polymer Solutions Workshop (Edinburgh, January 2025) and the EFDC2 (Dublin, August 2025), where the results of this study, in particular the proposition that the center mode by curving streamline enables the hoop stress mode to arise in straight pipes, were originally presented. Z.L. and B.H. acknowledge the technical support of ISTA Machine Shop, in particular Astrit Arslani, and ISTA Imaging and Optics Facility, in particular Robert Hauschild. The work was supported by the Austrian Science Fund (FWF) under grants I4188 and I06418, within the Deutsche Forschungsgemeinschaft research unit FOR 2688.

\bibliography{bib_elastic}

\end{document}